\documentclass[12pt]{book}

\usepackage[a4paper,margin=3cm,innermargin=3cm]{geometry}

\usepackage{multirow}
\usepackage{caption}
\usepackage{needspace}
\usepackage{marginnote}

\usepackage{imakeidx}
\usepackage{hyperref}
\makeindex[intoc]
\usepackage[table,xcdraw]{xcolor}

\newenvironment{conf-abstract}[4][]{
  \needspace{10\baselineskip}
  \begin{center}
    { \renewcommand\textsuperscript[1]{}
      \phantomsection\addcontentsline{toc}{section}
      {\texorpdfstring{#2 (\emph{#3})}{#2 (#3)}}
    }
    {{\large\bfseries #2}\marginnote{#1}\par}
    \medskip
    {#3\par}
    \smallskip
    {\small #4\par}
  \end{center}
}{%
  \bigskip
  \hrule
  \bigskip
}

\usepackage{etoolbox}

\usepackage{amsmath}
\usepackage{lipsum}

\def\bx{\boldsymbol{x}}

\def\bX{\boldsymbol{X}}

\def\bmu{\boldsymbol{\mu}}

\def\bSigma{\boldsymbol{\Sigma}}

\def\bPsi{\boldsymbol{\Psi}}

\def\g{\text{g}}
\def\r{{(r)}}
\def\rr{{(r + 1)}}

\begin{document}

%
%
%
%
%
%
%
%
%

\begin{conf-abstract}
{Cluster analysis and outlier detection with missing data}
{Hung Tong, Cristina Tortora}
{San Jos\'e State University}

A mixture of multivariate contaminated normal (MCN) distributions is a useful model-based clustering technique to accommodate data sets with mild outliers. However, this model only works when fitted to complete data sets, which is often not the case in real applications. In this paper, we develop a framework for fitting a mixture of MCN distributions to incomplete data sets, i.e. data sets with some values missing at random. We employ the expectation-conditional maximization algorithm for parameter estimation. We use a simulation study to compare the results of our model and a mixture of Student's $t$ distributions for incomplete data.

\section{Introduction}
Finite mixture models, or mixture models in short, assume that a population is a mixture of smaller sub-populations, each of which can be modeled by a probability distribution. The use of mixture models has been a powerful tool in cluster analysis, because each component of the mixture can represent a cluster. Moreover, basing cluster analysis on a mixture model enables formal inference on goodness of fit and choosing the number of clusters. Some mixture models, such as the Gaussian mixture model (GMM), can be appealing due to their computational and theoretical convenience. However, recent literature has moved away from the classic GMM; and many other distributions have been used to model the component densities, e.g. McNicholas (2016). One of the drawbacks of GMM is the lack of robustness when data are characterized by outliers; extreme points can affect the estimates of the cluster mean vectors and the covariance matrices. A natural solution is to use a robust component density, like the multivariate Student-$t$, or the contaminated normal (CN) distribution (Punzo and McNicholas, 2016). The use of those distributions would guarantee robust parameter estimation; moreover, the CN distribution can detect the outlying observations. It is important to highlight that both distributions can be used in the presence of mild outliers, i.e. outliers that are sampled from some populations different or even far from the assumed distribution, but cannot be used with data characterized by gross outliers, i.e. outliers that cannot be modeled by a distribution (Ritter, 2015, pp. 79–80). 
The aforementioned techniques have been shown to give strong classification performances on complete data sets, but they cannot work, as is, when fitted to data sets with missing values. The extension of existing clustering techniques for data with missing values has been the center of some field literature like Wang et al. (2004). The expectation-maximization (EM) algorithm (Dempster et al., 1977) used for parameter estimation for mixture models can be extended to be used with data sets with missing values. Following the idea of Wang et al. (2004) that focuses on mixtures of multivariate Student'$t$ distributions that can handle missing values, the goal of this paper is to extend the mixture of CN distribution for data sets with values missing at random.
\section{Background}
A $d$ dimensional random vector $\bX=\left(X_1,\ldots,X_d\right)^\top$ is said to follow a multivariate contaminated normal (MCN) distribution with mean vector $\bmu$, scale matrix $\bSigma$, proportion of good points {$\alpha\in\left(0,1\right)$}, and degree of contamination $\eta>1$ if its joint probability density function (pdf) is given by
\begin{align} 
f_{\text{MCN}}\left(\bx;\bmu,\bSigma,\alpha,\eta\right)&=\alpha f_{\text{MN}}\left(\bx;\bmu,\bSigma\right)+\left(1-\alpha\right)f_{\text{MN}}\left(\bx;\bmu,\eta\bSigma\right),
\label{eq:MCN distribution}
\end{align}
where $f_{\text{MN}}\left(\cdot;\bmu,\bSigma\right)$ denotes the pdf of a $d$-variate random vector that follows a multivariate normal (MN) distribution with mean vector $\bmu$ and covariance matrix $\bSigma$. The MCN distribution is a two-component mixture in which one component, with probability $\alpha$, represents the good observations, and the other component, with a probability $1 - \alpha$ represents the bad observations (or outliers). The two components share the same mean vector $\bmu$, but the component that represents the bad observations has an inflated covariance matrix $\eta \bSigma$. 
Two advantages of \eqref{eq:MCN distribution} are that once the parameters are estimated
we can establish if a generic point $\bx^*$ is good via the \textit{a~posteriori} probability, and that the estimation of $\bmu$ and $\bSigma$ are robust.

A $d$ dimensional random vector $\bX$ is said to follow a mixture of $G$ MCN (MCNM) distributions if its pdf can be written as
\begin{align} 
f_\text{MCNM} \left( \bx; \bPsi \right) = \sum_{\g = 1}^G \pi_\g f_\text{MCN} \left( \bx; \bmu_\g, \bSigma_\g, \alpha_\g, \eta_\g \right),\nonumber
\end{align}
\noindent where $\pi_\g$ is the mixing proportion of the gth component, such that $\pi_\g > 0$ and $\sum_{\g = 1}^G \pi_\g = 1$; the gth component is a MCN distribution as defined in \eqref{eq:MCN distribution},  $\bPsi = \left \{ \boldsymbol{\pi}, \boldsymbol{\vartheta} \right \}$, with $\boldsymbol{\pi} = \left \{ \pi_\g \right \}_{\g = 1}^G$, $\boldsymbol{\vartheta}_ = \left \{ \boldsymbol{\vartheta}_\g \right \}_{\g = 1}^G$, $\boldsymbol{\vartheta}_\g=\{\bmu_\g,\bSigma_\g,\alpha_\g,\eta_\g\}$.
The expectation-conditional maximization (ECM) algorithm (Meng and Rubin, 1993), a variant of the EM algorithm, is used for parameter estimation. The EM algorithm is based on the maximization of the complete-data likelihood, i.e. the
likelihood of the observed data $ \bx_i$ together with the unobserved
data. The algorithm iterates between two steps. In the E-step, the expected value of the complete-data likelihood is obtained. In the M-step, unknown parameters will be updated with those that maximize the expected value obtained in the E-step. When the unknown parameters cannot be updated independently from each other, the ECM algorithm is used, where the M-step is replaced by two simpler CM-steps.
\section{Methodology}
MCNM has two sources of missing data: component memberships $\boldsymbol{Z} = \left \{ \boldsymbol{z}_i \right \}_{i = 1}^n$, where $\boldsymbol{z}_i = \left( z_{i1}, \cdots, z_{iG} \right)^\top$ so that $z_{i \g} = 1$ if observation $i$ belongs to component $\g$, and $z_{i \g} = 0$ otherwise; and  whether observation $i$ is a good or bad point in each component $\boldsymbol{V} = \left \{ \boldsymbol{v}_i \right \}_{i = 1}^n$, where $\boldsymbol{v}_i = \left( v_{i1}, \cdots, v_{iG} \right)^\top$ so that $v_{i \g} = 1$ if observation $i$ is a good point in component $\g$, and $v_{i \g} = 0$ otherwise. When the data are characterized by missing values, there is a third source of missing data; each observation $\bx_i$ can be decomposed into $\left( \bx_i^o, \bx_i^m \right)$ where $\bx_i^o$ and $\bx_i^m$ denote the observed and missing values. Note that this notation does not imply the pattern of missingness is the same across all observations; for simplicity, we adopt $o$ and $m$ for the superscripts denoting sub-vectors rather than $o_i$ and $m_i$, which accurately represents how each observation can have a different number of missing values.

The complete-data likelihood of MCNM with missing values is given by $\mathcal{D} = \left \{ \bX^o, \bX^m, \boldsymbol{Z}, \boldsymbol{V} \right \} = \left \{ \bx_i^o, \bx_i^m, \boldsymbol{z}_i, \boldsymbol{v}_i \right \}_{i = 1}^n$, and the complete-data log-likelihood can be written as 
$
l (\bPsi; \mathcal{D}) = l(\boldsymbol{\pi}; \mathcal{D}) + l(\boldsymbol{\alpha}; \mathcal{D}) + l(\boldsymbol{\theta}; \mathcal{D})$ with
\begin{align*}
l(\boldsymbol{\pi}; \mathcal{D}) &= \sum_{i = 1}^n \sum_{\g = 1}^G z_{i \g} \ln \pi_\g, \hspace{0.6in} l(\boldsymbol{\alpha}; \mathcal{D}) = \sum_{i = 1}^n \sum_{\g = 1}^G z_{i \g} \left[ v_{i \g} \ln \alpha_\g + (1 - v_{i \g}) \ln (1 - \alpha_\g) \right], \\
l(\boldsymbol{\theta}; \mathcal{D}) &= -\cfrac{1}{2} \sum_{i = 1}^n \sum_{\g = 1}^G z_{i \g} \left \{ \ln | \bSigma_\g | + d_i^o (1 - v_{i \g}) \ln \eta_\g + \left( v_{i \g} + \cfrac{1 - v_{i \g}}{\eta_\g} \right) \delta \left(\begin{bmatrix} \bx^o_i \\[0.1in] \bx^m_i  \end{bmatrix}, \bmu_\g; \bSigma_\g \right) \right \},
\end{align*}
where $\delta (\cdot,\bx^m_i  , \bSigma_\g)$ is the squared Mahalanobis distance
and $d^o_i$ is the dimension of $\bx_i^o$. The ECM algorithm iterates between three steps, one E-step, and two CM-steps, until convergence. The E-step for the $(r + 1)-$th iteration requires the calculation of the following expectations
\begin{align*}
    & E_{\bPsi^\r} \left( Z_{i \g} \mid \bx_i^o \right)  =: \tilde{z}_{i \g}^\r, 
     & E_{\bPsi^\r} \left( X_i^m \mid \bx_i^o, Z_{i \g} = 1, V_{i \g} = 1 \right) =: \tilde{\bx}_{i \g}^\r, \\
    & E_{\bPsi^\r} \left( V_{i \g} \mid \bx_i^o, Z_{i \g} = 1 \right)  =: \tilde{v}_{i \g}^\r, 
    & E_{\bPsi^\r} \left( X_i^m X_i^{m, \top} \mid \bx_i^o, Z_{i \g} = 1, V_{i \g} = 1 \right)  =: \tilde{\tilde{\bx}}^\r_{i\g}.
\end{align*}
\noindent In calculating these expectations, it is important to recognize that the MCN distribution is a MN distribution itself, and if $\boldsymbol{Y}$ follows a MN distribution with mean vector $\bmu$ and covariance matrix $\bSigma$ such that 
\[
\boldsymbol{Y} = \begin{bmatrix} \boldsymbol{Y}_{(1)} \\ \boldsymbol{Y}_{(2)} \end{bmatrix}, \qquad
\bmu= \begin{bmatrix} \bmu_{(1)} \\[0.1in] \bmu_{(1)}  \end{bmatrix}, \qquad \bSigma = \begin{bmatrix} \bSigma_{11} & \bSigma_{12} \\[0.1in] \bSigma_{21} & \bSigma_{22} \end{bmatrix},
\]
where $\boldsymbol{Y}_{(1)}$ and $\bmu_{(1)}$ are $d_1$ dimensional vector, and $\bSigma_{11}$ is a $d_1 \times d_1$ matrix, then $\boldsymbol{Y}_{(1)}$ follows a MN distribution with mean vector $\bmu_{(1)}$ and covariance matrix $\bSigma_{11}$.

The first CM-step holds $\boldsymbol{\eta}$ fixed at $\boldsymbol{\eta}^\r$ and updates $\boldsymbol{\pi}, \boldsymbol{\alpha}, \bmu, \bSigma$ as followed
\[
\pi^\rr_\g = \cfrac{1}{n}\sum_{i = 1}^n \tilde{z}^\r_{i\g}, \hspace{0.7in} \alpha^\rr_{\g}=\cfrac{\sum_{i=1}^n \tilde{z}^\r_{i\g}\tilde{v}^\r_{i\g}}{\sum_{i = 1}^n \tilde{z}^\r_{i\g}},
\hspace{0.7in} 
\tilde{w}^\r_{i\g}=\tilde{z}^\r_{i\g} \left( \tilde{v}^\r_{i\g} + \cfrac{1 - \tilde{v}^\r_{i\g}}{\eta^\r_\g}\right),
\]
\vspace{-0.1in}
\[
\bmu^\rr_{\g}=\cfrac{\sum_{i=1}^n \tilde{w}^\r_{i\g}\begin{bmatrix} \bx^o_i \\[0.1in] \tilde{\bx}^\r_{i\g} \end{bmatrix}}{\sum_{i = 1}^n \tilde{w}^\r_{i\g}}, \hspace{0.4in} \bSigma^\rr_\g=\cfrac{\sum_{i=1}^n\tilde{w}^\r_{i\g}
\tilde{\bSigma}^\r_{i\g}}{\sum_{i=1}^n \tilde{z}^\r_{i\g}}, 
\]
\noindent  $\tilde{\bSigma}^\r_{i\g} = \begin{bmatrix}
\left( \bx^o_i - \bmu^{\r, o}_\g \right)\left( \bx^o_i - \bmu^{\r, o}_\g \right)^\top & \left( \bx^o_i - \bmu^{\r, o}_\g \right)\left(\tilde{\bx}^\r_{i\g} - \bmu^{\r, m}_\g \right)^\top \\[0.2in]
\left(\tilde{\bx}^\r_{i\g} - \bmu^{\r, m}_\g \right)\left( \bx^o_i - \bmu^{\r, o}_\g \right)^\top & \qquad \left(\tilde{\bx}^\r_{i\g} - \bmu^{\r, m}_\g \right)\left(\tilde{\bx}^\r_{i\g} - \bmu^{\r, m}_\g \right)^\top + \tilde{\tilde{\bx}}^\r_{i\g} - \tilde{\bx}_{i \g}^\r \ \tilde{\bx}_{i \g}^{\r, \top} 
\end{bmatrix}$.

The second CM-step  updates $\boldsymbol{\eta}$ as followed
\[
\eta^\rr_\g = \cfrac{\sum_{i=1}^n \tilde{z}^\r_{i\g} \left( 1 - \tilde{v}^\r_{i\g} \right) \delta \left(\begin{bmatrix} \bx^o_i \\[0.1in] \tilde{\bx}_{i \g}^\r  \end{bmatrix}, \boldsymbol{\mu}^\rr_\g; \boldsymbol{\Sigma}^\rr_\g \right)}{ \sum_{i=1}^n d_i^o \ \tilde{z}^\r_{i\g} \left( 1 - \tilde{v}^\r_{i\g} \right)}.
\]
The E-step and the CM-steps are iterated until convergence is reached.
\section{Application}
A simulation study similar to the one conducted by Punzo and McNicholas (2016) was performed to compare the results of the MCNM and mixture of Student' $t$ distributions with missing data. We fixed $G= 2$, and we generated data from different scenarios, varying the number of observations, 100 or 500, and the level of cluster overlap, far or close. Moreover, clusters were generated from one of the following four bivariate two-component mixtures: (a) Student's $t$, (b) MCN, (c) MN with $1\%$ of points randomly substituted by high atypical points, and (d) MN with $5\%$ of points randomly substituted by noise points generated from a uniform distribution. We then hid some values of $10\%$, $50\%$ or $80\%$ observations, for a total of $48$ scenarios. When hiding values, we applied $10$ general missing patterns proposed by \texttt{ampute} of the \textbf{mice} package. Under each scenarios, we generated $20$ data sets, for a total of $9,600$ simulations. For each simulation, we recorded the adjusted Rand indices (ARI) produced by the MCNM and mixture of Student'$t$ distributions. For those that involve (c) and (d), we also recorded their true positive rates (TPR) and false positive rates (FPR) in outlier detection.  On average, the results indicated a comparable performance of the two mixtures in term of ARI, but better performance of the MCNM in term of FPR. Moreover, the ARIs decrease for higher proportions of observations with missing values, which is expected.
\vspace{-.1cm}
\section{References}
Buuren S. and Groothuis-Oudshoorn K. (2011). \textbf{mice}: Multivariate Imputation by Chained Equations in \textsf{R}. {\em Journal of Statistical Software}, {\bf 45}(3), 1-67. 
\medskip \\ 
McNicholas, P.D. (2016).  Mixture Model-Based Classification. {\em Boca Raton FL: Chapman \& Hall/CRC}.
\medskip \\
Meng, X., and Rubin, D. (1993). Maximum Likelihood Estimation via the ECM Algorithm: A General Framework. {\em Biometrika}, {\bf 80}(2), 267-278.
\medskip \\
Punzo, A., and McNicholas, P. D. (2016). Parsimonious Mixtures of Multivariate Contaminated Normal Distributions. {\em Biometrical Journal}, {\bf 58}(6), 1506-1537.
\medskip \\
Ritter, G. (2015). Robust Cluster Analysis and Variable Selection. {\em New York: Chapman \& Hall/CRC}.
\medskip \\
Wang, H., Zhang, Q., Luo, B., and Wei, S. (2004). Robust mixture modelling using multivariate t-dist. with missing information. {\em Pat. Rec. Let.}, {\bf 25}(6), 701--710.
\end{conf-abstract}



\end{document}